\documentstyle[fleqn]{tp}

\input psfig

\def\kms{\mbox{km s$^{-1}$}}

\def\la      {\mathrel{\hbox{\rlap{\hbox{\lower4pt\hbox{$\sim$}}}\hbox{$<$}}}}
\def\ga      {\mathrel{\hbox{\rlap{\hbox{\lower4pt\hbox{$\sim$}}}\hbox{$>$}}}}

\begin{document}

\twocolumn[
\title{Antibiasing: High mass-to-light ratios in dense clusters}
\author{R. Brent Tully$^1$ and Edward J Shaya$^2$\\
{\it $^1$Institute for Astronomy, University of Hawaii}\\ 
{\it 2680 Woodlawn Dr., Honolulu, Hawaii, 96822, USA}\\
{\it $^2$Hughes STX, Goddard Space Flight Center, Code 631}\\ 
{\it Greenbelt, Maryland 20771, USA}}
\vspace*{16pt}   

ABSTRACT.\
Modeling of the velocity field of the Local Supercluster leads to the
conclusion that there must be a substantial differential between the 
mass-to-light ratios of most galaxies and those in a small number of special
places.  Specifically, $M/L=1000 M_{\odot}/L_{\odot}$ is indicated for
the full Virgo Cluster while $M/L=150 M_{\odot}/L_{\odot}$ is appropriate
for most environments.  It is argued that a higher $M/L$ value is
characteristic of E/S0 knots, regions where the galaxy density is high
and collapse time scales are short.  Regions larger than galaxies where
collapse is ocurring only on the order of the age of the universe, the
environment where spirals and irregulars predominate, are characterized by
a low $M/L$.  Overall, the mean density of the universe is well below
closure.  There could be comparable total mass in the relatively rare
E/S0 knots as there is in the environments dominated by spirals and
irregulars.
\endabstract]

\markboth{R. Brent Tully \& Edward J. Shaya}{Antibiasing}

\small

\section{Introduction}

The concept of `biasing' arose out of a discussion by Kaiser (1984) that
different classes of astronomical objects may trace the underlying
mass distribution to greater or lesser degrees.  It was appreciated,
especially as comparisons were made with N-body simulations, that some
fraction of mass could be only poorly correlated with the observable
galaxies.  A simple, perhaps simplistic, description is provided by the
bias parameter of linear theory, $b = \delta_g / \delta_m$, where 
the {\it observed} galaxy fluctuation field is $\delta_g$ and the mass density
fluctuation field is $\delta_m$.  The expectation has arisen from comparisons
between models and the observed spectrum of irregularities (Davis et al. 1985)
that mass would
be more widely distributed than clusters and even field galaxies, so
$b>1$.  In this case, it would be said that galaxies have a biased distribution
compared with the mass.

While this sense of biasing may well occur, the focus of this paper is on
the opposite possibility that matter may be more concentrated than light
in some cosmologically important situations.  At least locally, it could
be that $b<1$, a condition we will call `antibiasing'.

\section{The Mean Density of the Universe}

Dynamical studies of galaxy flows can give estimates of the mean density
(Dekel 1994; Strauss \& Willick 1995).  There is a dichotomy of results.
One method compares the divergence of the velocity field, which gives the 
mass density field, with some catalog of the observed distribution of galaxies.
This method, called POTENT, gives values of $\beta = \Omega_0^{0.6}/b$
that are indicative of a universe near closure density:
$\beta_I=0.89\pm0.24$ (Sigad et al. 1998) and $\beta_O=0.74\pm0.26$
(Hudson et al. 1995) where the subscripts $I$ and $O$ indicate comparison 
with an Infrared Astronomy Satellite (IRAS) and optical catalog, respectively.
Baker et al. (1998) find $b_O/b_I \simeq 1.4$, so these results would be
compatible with $\Omega_0 \sim 1$ and $b_I \sim 1$. The uncertainties
are 95\% confidence levels.

On the other hand, methods that compare expectation and observed velocities
find lower values of $\beta$.  Here, both linear and non-linear studies
reach similar conclusions.  The linear approximation is valid if 
$\delta_m \la 1$ (Peebles 1980):
$$
{\bf v} = {\beta \over 4\pi} \int d^3{\bf r^{\prime}} 
{\delta_g({\bf r^{\prime}})({\bf r^{\prime}}-{\bf r}) \over 
\mid {\bf r^{\prime}}-{\bf r} \mid^3.} 
$$
Some results: Shaya, Tully, \& Pierce (1992) found $\beta_O \la 0.25$ 
($\beta_I \la 0.35$), Reiss et al. (1997) found $\beta_I=0.4_{-0.25}^{+0.3}$,
da Costa et al. (1998) found $\beta_I=0.6\pm0.2$, and Willick \& Strauss
(1998) found $\beta_I=0.50\pm0.10$.  A non-linear methodology we call
`Least Action' has been described by Peebles (1989, 1990, 1994, 1995)
and Shaya, Peebles \& Tully (1995).  
The non-linear studies determine, not $\beta$, but constraints on the 
parameter space ($M/L$, $t_0$), where $M/L$ is the conversion ratio from
observed luminosities to mass and $t_0$ is the age of the universe.  The 
$M/L$ values that are determined can be normalized by the mean luminosity
density of the universe to constrain something equivalent to $\beta$.  
The results from our most recent analysis
(Tully 1997) can be translated as $\beta_O=0.31\pm0.14$ 
($\beta_I=0.43\pm0.20$).  All the linear and non-linear studies that 
compare model velocities to observed velocities are in reasonable
agreement and indicate that if $\Omega_0=1$ then there needs to be 
considerable bias, $b_I \sim 2$.

\section{Least Action}

Shaya et al. (1995) describe our methodology for the reconstruction of
the orbits of galaxies and groups using a variational principle.
Tully (1997) gives updated results.  Galaxies follow paths that extremize
the `action', the integral of the Lagrangian through time.  In its
simplest incarnation, the model can be restricted to only two free
parameters.  We have a catalog of galaxies within a velocity limit
(3000~\kms) with known positions on the sky, redshifts, and luminosities,
so one of our free parameters, the relation between mass and light
($M/L$), specifies the mass associated with galaxies where they are today
in redshift space.  The other free parameter is the age of the universe,
$t_0$, that specifies how long these masses have been tugging on each other.
At the next level of complexity, a third parameter is added to take some
account of biasing.  A softening parameter is introduced to modify the 
force law so forces are reduced at close range to mimic the effects of
inter-penetrating halos.

Once the three free parameters have been fixed, a Least Action re-enactment
of the constituent orbits gets all the galaxies to the right positions on 
the sky and the right redshifts (as boundary conditions) and specifies
the other three elements of phase-space.  The specification of model
distances is what we care about because measured distances exist for
$\sim 1/3$ of the objects.  A $\chi^2$ evaluator is formulated to compare
model and observed distance moduli, $\mu _{m,i}$ and $\mu _{o,i}$
respectively, for the $i^{th}$ object:
$$
\chi^2 = \sum_{i=1}^{N_{\mu}} w_i \Big(({{\mu_{o,i}-\mu_{m,i}} \over \sigma_{\mu}})^2
$$
$$
        +  ({{5 log (cz_{o,i} / cz_{m,i})} \over \sigma_{\mu}})^2\Big)
        / \sum_{i=1}^{N_{\mu}} w_i
$$

Figure~1 illustrates the $\chi^2$ values in the domain of the two principal
free parameters ($M/L$, $t_0$) with the softening parameter fixed at 300~\kms.
(Note: since this model was created we have revised our table of distances;
the Hubble Space Telescope observations of cepheids in external galaxies
has resulted in a 10\% upward revision of distances and ages: see Tully 1998).
It can be seen that models with $\Omega_0$ substantially less than unity are
favored. Models with $M/L\sim200 M_{\odot}/L_{\odot}$, $t_0\sim 10$~Gyr
conform to observations.

\begin{figure*}
\psfig{figure=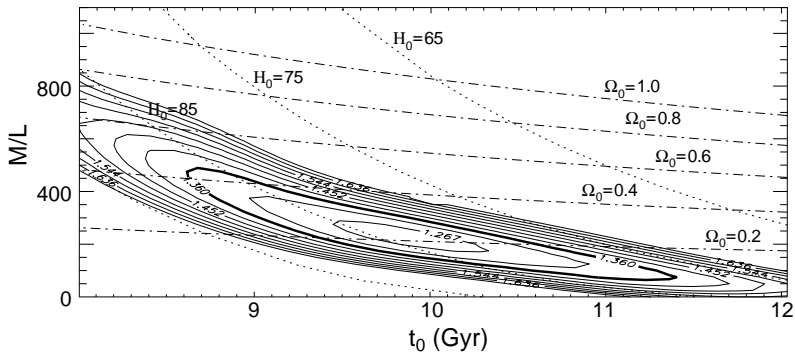,height=3.3in,rheight=2.in,bbllx=80bp,bblly=140bp,bburx=490bp,bbury=500bp}
\caption[]{Contours of $\chi^2$ as a function of the two free parameters
$M/L$ and $t_0$ with the force softening parameter set at 300~\kms.  Contours
at $1\sigma$ intervals with the dark contour at the $2\sigma$ level.
This model is calibrated with distances that do not reflect a 10\% increase
in the zero-point resulting from recent cepheid distance determinations.
}
\label{fig:1}
\end{figure*}

\section{The Virgo Cluster}

Although the model described above provides a good description of the 
velocity field over most of the Local Supercluster, it fails miserably
in one important region.  A substantial infall pattern is observed around 
the Virgo Cluster but such a low value as $M/L\sim200 M_{\odot}/L_{\odot}$
cannot explain the motions that are seen.  It hardly helps to simply increase
$M/L$ for all objects.  Doing so effectively increases the model $\Omega_0$,
hence reduces $t_0$ for a given distance scaling.  There is then not enough
time to build up the observed velocities.  High global $M/L$  cannot explain
the infall velocities {\it and} the $\chi^2$ figure of merit blows up.

The very strong conclusion is that we need a large {\it differential} in 
$M/L$ between Virgo and most of the rest of the Local Supercluster.  We need
a modest $M/L$ outside Virgo to get a good $\chi^2$ fit overall but we need a 
large $M/L$ in the Virgo Cluster to explain the observed infall pattern.

Figure~2 illustrates the nature of our observational constraints.  The group 
11-4 (Tully 1987) is an entity in the Southern Extension of the Virgo Cluster,
at an angle of $9.3^{\circ}$ from the cluster center.  The curves in each 
panel show the run of velocities with distance anticipated by two alternative
models: on top, a model with $\Omega_0=0.3$, $M/L=200 M_{\odot}/L_{\odot}$
and, on the bottom, a model with $\Omega_0=0.3$, $M/L=150 M_{\odot}/L_{\odot}$
for most galaxies but $M/L=1000 M_{\odot}/L_{\odot}$ for Virgo and a small
number of other groups dominated by ellipticals and S0s.  The points with
error bars locate galaxies in group 11-4 with distance estimates.  Clearly,
the model that defines the curve in the top panel cannot explain the 
velocities that are observed.  The model associated with the bottom panel 
does an adequate job.

\begin{figure*}
\centering\mbox{\psfig{figure=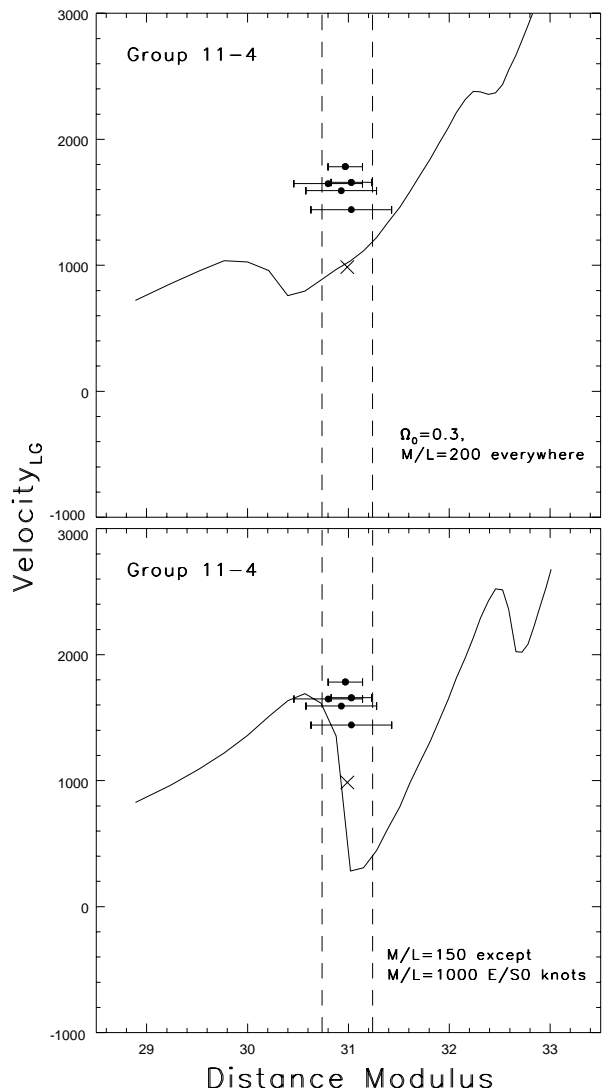,height=15cm}}
\vspace*{.1cm}
\caption[]{Example of model velocities along a line-of-sight close to the 
direction of the Virgo Cluster.  In this case, the sight line is through
Group 11-4.  The points with errors correspond to galaxies in this group
with distance determinations.  The vertical dashed lines bracket the
distance of the Virgo Cluster, centered in distance and velocity at the 
large cross.  Top panel: $M/L=200$ for all entities.  Bottom panel:
$M/L=1000$ for Virgo and other E/S0 knots, otherwise $M/L=150$.  The second
wave in the velocity curve beyond Virgo occurs because the line-of-sight
passes near another E/S0 knot, the Virgo~W Cluster, Group 11-24.
}
\label{fig:2}
\end{figure*}

All galaxies within the infall region can be considered in the same light.
In each case, the run of velocities with distance along the pertinent
line-of-sight can be calculated for any model.  For a given $t_0$, there will
be a threshold $(M/L)_{minimum}$ that allows the observed velocity within the 
infall envelope.

It is helpful to appreciate that we do not need very precise distance
estimates for this analysis.  It is sufficient to discriminate
between three possibilities: is a galaxy to the foreground, within the
infall region, or in the background?  Since the Virgo multi-value region
is seen to extend to $\sim25^{\circ}$ from the Virgo core, the infall
region extends to $\sim1.2^m$ in front of the mean cluster distance and to 
$\sim0.8^m$ behind the mean distance.  Our distance estimators are adequate
to discriminate between the three alternatives.  The case of Fig.~2 
provides an illustration.

\begin{figure*}
\centering\mbox{\psfig{figure=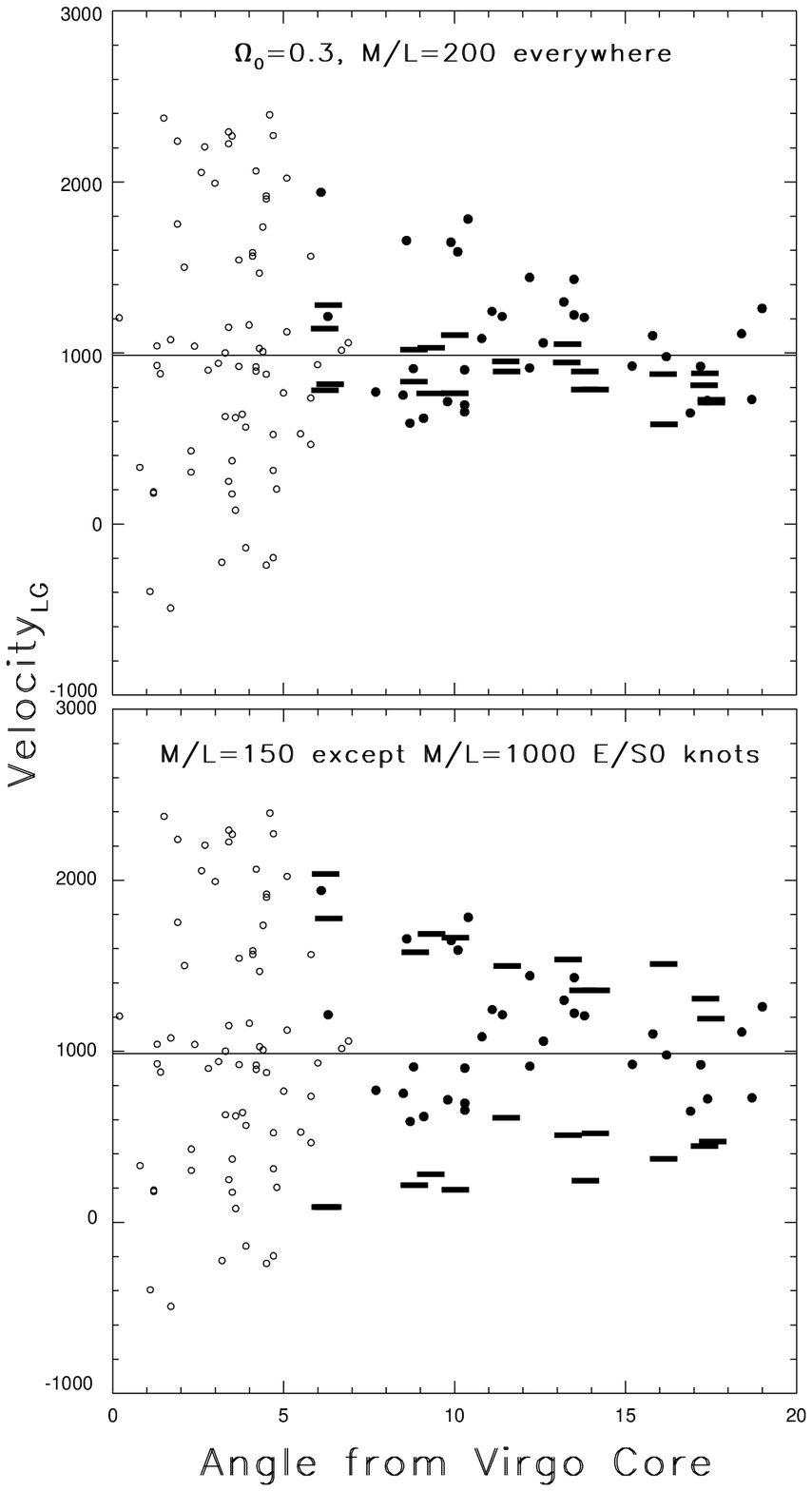,height=15cm}}
\vspace*{.1cm}
\caption[]{Virgo infall constraints from two models.  Horizontal bars are
line-of-sight extrema velocities for a good overall $M/L$ (top) and for a
model with a lot of extra mass in Virgo (bottom).
}
\label{fig:3}
\end{figure*}

The Virgo Cluster caustic lies at $\sim 6^{\circ}$ from the core and the
infall is pronounced out to $\sim 20^{\circ}$.  Within this annulus, we 
currently have 35 galaxies with distance determinations that unambigously
put them in the infall zone.  These galaxies are falling in a quasi-laminar
flow into the cluster.  Figure~3 shows the constraints these galaxies provide 
for the same two models considered in Fig.~2.  The circles show the 
observed velocities and angular distances from the cluster center
of galaxies that are certified to be within the Virgo infall zone on the basis
of distance measurements.  The small open circles correspond to galaxies 
projected onto the cluster and, hence, are probable cluster members.  The
35 filled circles correspond to galaxies projected outside the cluster
proper.  The horizontal bars in each panel indicate the maximum and minimum
projected velocities along lines of sight to individual galaxies or groups 
of galaxies within the infall region.  The 11-4 group used as 
an example above is at an angle of $9.3^{\circ}$ and the maximum and minimum
velocities anticipated by the single value $M/L=200 M_{\odot}/L_{\odot}$
model (top panels Figs. 2 and 3) are 1037 and 760~\kms, while the maximum
and minimum velocities anticipated by the model with 
$M/L=1000 M_{\odot}/L_{\odot}$
for Virgo, $M/L=150 M_{\odot}/L_{\odot}$ for most other objects (bottom
panels Figs. 2 and 3) are 1690 and 283~\kms.  The model illustrated in the
top panels totally fails to predict the observed infall pattern while the
model displayed in the bottom panels is adequate.  In this case,
the Virgo Cluster is assigned $1.3\times10^{15} M_{\odot}$.  This mass
is higher than the Virial theorem mass of
$0.7\times10^{15} M_{\odot}$ (Tully \& Shaya 1984), but the infall mass
pertains to a scale $\sim 3\times$ larger than the Virial radius.

Fifteen years ago, Tully \& Shaya (1984) had already found this basic result
that $(M/L)_{virgo}$ has to be much higher than $(M/L)_{field}$ in order to get
a sensible infall description.  That previous study was based on a fully 
non-linear but spherically symmetric model and one could wonder if
the obvious departure of the observed galaxies from a spherical distribution
in the Local Supercluster was affecting the solution.  The present model 
provides
a non-parametric description of the supercluster asymmetry (mass is distributed
like the observed galaxies) and returns essentially the same answer.  The
infalling galaxies in Fig.~3 tend to have higher velocities than the cluster 
mean, partially a happenstance of uneven filling of the volume and partly
due to angular momentum in the infalling population, naturally provided by
the Least Action reconstruction.  This velocity asymmetry was seen in the 1984 
study but could not be described by the spherically symmetric model
(see also Tully \& Shaya 1994; Tully 1997).

\section{High $M/L$ in a Small E/S0 Knot}

Throughout the Local Supercluster there are places where the galaxy density
is as high as in the core of the Virgo Cluster, though the total number of
galaxies involved can be small.  Usually in these cases the majority of the
galaxies are ellipticals and S0s.  There is some evidence that $M/L$ values
can be very high in these places.  A dramatic case is the group around
NGC~1407, Group 51-8 in our catalog (Tully 1987).  Its unusual properties
have been discussed by Gould (1993) and Quintana, Fouqu\'e, \& Way (1994).
There are 19 galaxies with known redshifts associated with the group.
Only NGC~1407 is an $L^\star$ galaxy.  The second most luminous galaxy,
NGC~1400, is more than three times fainter.  All but one of the dozen that
have been typed are E/S0/Sa.  Quintana et al. find the velocity dispersion
to be $416_{-58}^{+98}$~\kms, which gives 
$(M/L)_{virial}=600 M_{\odot}/L_{\odot}$
after a tiny correction to our distance scale.

The S0 galaxy NGC~1400 deserves special attention.  This galaxy is 
blueshifted by 1066~\kms\ with respect to the group mean and would carry
the vast majority of the kinetic energy of the group ($\sim 85\%$ of the
kinetic energy follows from the line-of-sight velocities).  One way to 
rationalize this unusual situation is to suppose that NGC~1400 and the
other galaxies carry only a small fraction of the mass of the gravitational 
potential well.

\begin{table*}
\caption[]{E/S0 Knots within 3000~\kms  }
  \centering
  \begin{tabular}{lcrcc}
    \hline
    Name & Catalog & $V_{LG}$ & Log $L_B$ & \\
    \hline\hline
    Virgo         & 11~~-1 &  986 & 12.14 & $\ast$ \\
    Fornax        & 51~~-1 & 1466 & 11.52 & $\ast$ \\
    Virgo~W       & 11-24  & 2225 & 11.43 & \\
    Coma~I        & 14~~-1 &  983 & 11.34 & $\ast$ \\
    NGC~5044      & 11-31  & 2559 & 11.19 & \\
    Antlia        & 31~~-2 & 2519 & 11.12 & \\
    Eridanus      & 51~~-4 & 1415 & 11.06 & $\ast$ \\
    NGC~1566      & 53~~-1 & 1076 & 11.03 & $\ast$ \\
    NGC~4125      & 12~~-5 & 1436 & 11.00 & $\ast$ \\
    NGC~5846      & 41~~-1 & 1810 & 10.99 & \\
    Leo           & 15~~-1 &  722 & 10.73 & $\ast$ \\
    NGC~1407      & 51~~-8 & 1522 & 10.58 & $\ast$ \\
    \hline
  \end{tabular}
\label{tab:1}
\end{table*}

\section{Other E/S0 Knots in the Local Supercluster}

In a complete volume-limited sample in our vicinity of the Local Supercluster
(extending to $25 h_{75}^{-1}$~Mpc with distances given by a model with
spherical Virgo infall), 69\% of the galaxies, and 77\% of the light in 
galaxies, are in 179 groups (Tully 1987).  Spiral and irregular galaxies
predominate in the overwhelming majority of these groups.  In these most
common cases, the E/S0 representation is inevitably less than 20\%,
1-dimensional group velocity dispersions are $\sim 100$~\kms, and group
crossing-times are $\sim 40\%$ of the Hubble time.  However,
eight of these groups are quite distinct.  Indeed, they are among the best 
known nearby groups because of their compact nature.  In these
eight cases, {\it more than half of the galaxies are E/S0.}  These early
type galaxies are inevitably restricted to volumes only a few hundred kpc
across.  The groups usually have higher velocity dispersions than their 
spiral-dominated counterparts and crossing times are $\la 10\%$ of the 
Hubble time.  The distinction between spiral-rich and E/S0-rich groups is
quite pronounced, obscured only partially by the not-infrequent incidence
of spirals at the periphery of E/S0 knots.

Table~1 provides a list of the complete sample of nearby groups 
dominated by elliptical and S0 galaxies
(the asterisk in the final column indicates inclusion in the complete sample), 
plus a few other such groups at greater distances but still within 3000~\kms.
The most luminous and the faintest groups on the list have already been
discussed.
The Virgo Cluster and the NGC~1407 Group span a range from two dozen
$L^{\star}$ galaxies and $10^{15} M_{\odot}$ to a single $L^{\star}$ galaxy 
and $2\times10^{13} M_{\odot}$.  Fornax, Coma~I, Antlia, and NGC~1566
all have internal velocity dispersions that hint at large $M/L$ values
(Tully 1987).

\section{Summary}

The evidence for large $M/L$ variations comes from combining two distinct
sources of dynamical information.  On the one hand, the Least Action
reconstruction of orbits is telling us that most of the galaxies
(ie, the spirals and irregulars) live in an environment with a modest mean
density. On the other hand, the large infall motions into the Virgo Cluster
and the large internal dispersions in other E/S0 knots are indications
of a lot of mass in these special locations.  The Least Action dynamical
measurements encompass larger scales than virial measurements of internal
group motions which probably explains why the Least Action mass estimates
are larger by up to a factor 2.  It is suggested that 
$M/L\sim150 M_{\odot}/L_{\odot}$ for most galaxy associations but
$M/L\sim1000 M_{\odot}/L_{\odot}$ in the knots of ellipticals and S0s.
Weak lensing provides a probe of
the distribution of matter on supercluster scales for distant objects and it
is interesting that hints are arising from that work that mass is strongly
`antibiased' toward the clusters (Kaiser et al. 1998).

If E/S0 knots do have much higher $M/L$ values than galaxies in lower
density regions there are several potential astrophysical explanations:
(i) dimming at $L_B$ is anticipated for older stellar populations, 
(ii) stripping from encounters might shred a large fraction of stars 
into intracluster space; Mendez et al. (1997) and Ferguson et al. (1998)
find enough intracluster planetary nebulae and red giants between the 
galaxies in Virgo to indicate that there are comparable numbers of stars 
outside galaxies as in, (iii) dwarf spheroidal galaxies are found in large 
numbers in Virgo (Phillipps et al. 1998) and Fornax (Ferguson \& Sandage 1988) 
but evidently {\it not} in large numbers in
Ursa Major or the field (Trentham, Tully, \& Verheijen, in preparation) 
(iv) field galaxies may continue to
accrete gas even until today, gas that gets converted into stars, while
gas that arrives in clusters after the initial collapse of the cluster is
outside the Roche-limit of individual galaxies and becomes virialized as hot
gas in the intracluster medium (Shaya \& Tully 1984).  It is easy to entertain
that most or all these effects occur and cummulatively could account for an
$M/L$ differential of the observed factor of 6-7.

Roughly 8\% of the $B$ light within 3000~\kms comes from the galaxies
associated with E/S0 knots.  However, with the $M/L$ differential that has 
been suggested, the E/S0 knots could contain $\sim 40\%$ of the mass
directly associated with galaxies.  If so, there is a dramatic 
{\it antibiasing} of matter compared with IRAS-selected galaxies or even
optically selected galaxies.  Dark matter is more concentrated in a few
places but enough so that it would have a cosmological impact.  In
addition, there may also be the {\it bias} of some matter poorly 
correlated with galaxies.  The Least Action modeling strongly suggests 
that the overall density of matter is modest, $\Omega_0\sim0.25$.
The POTENT observations of $\Omega_0=(b \beta)^{1.7}$ can be reconciled 
with a value of $\Omega_0$ well less than unity if, overall, $b_I<1$.

%

\section*{Acknowledgments}

We thank our close collaborator in this research, Jim Peebles.



\begin{thebibliography}{99}

\bibitem{ref:1} 
Baker, J.E., Davis, M., Strauss, M.A., Lahav, O., \& Santiago, B.X.,
1998, astro-ph/9802173

\bibitem{ref:1} 
da Costa, L.N., Nusser, A., Freudling, W., Giovanelli, R., Haynes, M.P.,
Salzer, J.J., \& Wegner, G., 1998, MNRAS, 299, 425

\bibitem{ref:1} 
Davis, M., Efstathiou, G,. Frenk, C.S., \& White, S.D.M., 1985,
ApJ, 292, 371

\bibitem{ref:1} 
Dekel, A.,  1994, Ann. Rev. Astron. Astrophys., 32, 371

\bibitem{ref:1} 
Ferguson, H. C., \& Sandage, A., 1988, AJ, 96, 1520

\bibitem{ref:1} 
Ferguson, H.C., Tanvir, N.R., \& von Hippel, T.,  1998, Nature, 391, 461

\bibitem{ref:1} 
Gould, A., 1993, ApJ, 403, 37

\bibitem{ref:1} 
Hudson, M.J., Dekel, A., Courteau, S., Faber, S.M., \& Willick, J.A.,
1995, MNRAS, 274, 305

\bibitem{ref:2} 
Kaiser, N., 1984, ApJ, 284, L9

\bibitem{ref:1} 
Kaiser, N., Wilson, G., Luppino, G., Kofman, L., Gioia, I., Metzger, M.,
\& Dahle, H.,  1998, astro-ph/9809268

\bibitem{ref:1} 
Mendez, R.H., Guerrero, M.A., Freeman, K.C., Arnaboldi, M. et al., 
ApJ, 491, L23

\bibitem{ref:1} 
Peebles, P.J.E.,  1980, The Large-Scale Structure of the
Universe. Princeton University Press, Princeton

\bibitem{ref:1} 
Peebles, P.J.E.,  1989, ApJ, 344, L53

\bibitem{ref:1} 
Peebles, P.J.E.,  1990, ApJ, 362, 1

\bibitem{ref:1} 
Peebles, P.J.E.,  1994, ApJ, 429, 43

\bibitem{ref:1} 
Peebles, P.J.E.,  1995, ApJ, 449, 52

\bibitem{ref:1} 
Phillipps, S., Parker, Q.A., Schwartzenberg, J.M., \& Jones, J.B.,  1998,
ApJ, 493, L59

\bibitem{ref:1} 
Quintana, H., Fouqu\'e, P., \& Way, M.J.,  1994, A\&A, 283, 722

\bibitem{ref:1} 
Riess, A.G., Davis, M., Baker, J., \& Kirshner, R.P.,  1997, ApJ, 488, L1

\bibitem{ref:1} 
Sigad, Y., Eldar, A., Dekel, A,, Strauss, M.A., \& Yahil, A.,  1998,
ApJ, 495, 516

\bibitem{ref:1} 
Shaya, E.J., Peebles, P.J.E., \& Tully, R.B.,  1995, ApJ, 454, 15

\bibitem{ref:1} 
Shaya, E.J., \& Tully, R.B.,  1984, ApJ, 281, 56

\bibitem{ref:1} 
Shaya, E.J., Tully, R.B., \& Pierce, M.J.,  1992, ApJ, 391, 16

\bibitem{ref:1} 
Strauss, M.A., \& Willick, J.,  1995, {\it Physics Reports}, 261, 271

\bibitem{ref:1} 
Tully, R.B.,  1987, ApJ, 321, 280

\bibitem{ref:1} 
Tully, R.B., 1997, in Sato, K. ed., IAU Symp. 183, Cosmological Parameters and
Evolution of the Universe. Kluwer Acad. Publ., Dordrecht

\bibitem{ref:1} 
Tully, R.B. 1998, in Caputo, F. and Heck, A. eds., Post-Hipparcos Cosmic 
Candles. Kluwer Acad. Publ., Dordrecht. astro-ph/9809394

\bibitem{ref:1} 
Tully, R.B., \& Shaya, E.J.,  1984, ApJ, 281, 31

\bibitem{ref:1} 
Tully, R.B., \& Shaya, E.J.,  1994, in Durret, F., Measure, A., \&
Tran Thanh Van, J. eds., Clusters of Galaxies: 
Rencontre de Moriond. Editions Frontieres, Gif-sur-Yvette, p53

\bibitem{ref:1} 
Willick, J.A., \& Strauss, M.A.,  1998, astro-ph/9801307



\end{thebibliography}
\end{document}